\documentclass[3p]{elsarticle}
\usepackage{times}
\usepackage{epsfig}
\usepackage{setspace}
\usepackage{array}
\usepackage{multirow}
\usepackage{booktabs}
\usepackage{color}
\usepackage{amsfonts}
\usepackage{threeparttable}
\usepackage{amsmath,bm}
\usepackage{lscape}
\usepackage[pagebackref=true,breaklinks=true,colorlinks,bookmarks=false]{hyperref}
\DeclareMathOperator*{\argmin}{arg\,min} 
\DeclareMathOperator*{\argmax}{arg\,max}
\newcommand{\tabincell}[2]{\begin{tabular}{@{}#1@{}}#2\end{tabular}}
\journal{Journal of \LaTeX\ Templates}

\begin{document}

\begin{frontmatter}

  \title{Free-form tumor synthesis in computed tomography images via richer generative adversarial network}

  \author[mymainaddress,mythirdaddress]{Qiangguo~Jin}

  \author[mysecondaryaddress]{Hui~Cui}

  \author[mythirdaddress]{Changming~Sun}

  \author[mymainaddress,myfourthaddress]{Zhaopeng~Meng}

  \author[mymainaddress]{Ran Su}

  \address[mymainaddress]{School of Computer Software, College of Intelligence and Computing, Tianjin University, Tianjin, China}
  \address[mysecondaryaddress]{Department of Computer Science and Information Technology, La Trobe University, Melbourne, Australia}
  \address[mythirdaddress]{CSIRO Data61, Sydney, Australia}
  \address[myfourthaddress]{Tianjin University of Traditional Chinese Medicine, Tianjin, China}

  \begin{abstract}
    The insufficiency of annotated medical imaging scans for cancer makes it challenging to train and validate data-hungry deep learning models in precision oncology. We propose a new richer generative adversarial network for free-form 3D tumor/lesion synthesis in computed tomography (CT) images. The network is composed of a new richer convolutional feature enhanced dilated-gated generator (RicherDG) and a hybrid loss function. The RicherDG has dilated-gated convolution layers to enable tumor-painting and to enlarge perceptive fields; and it has a novel richer convolutional feature association branch to recover multi-scale convolutional features especially from uncertain boundaries between tumor and surrounding healthy tissues. The hybrid loss function, which consists of a diverse range of losses, is designed to aggregate complementary information to improve optimization.
    We perform a comprehensive evaluation of the synthesis results on a wide range of public CT image datasets covering the liver, kidney tumors, and lung nodules. The qualitative and quantitative evaluations and ablation study demonstrated improved synthesizing results over advanced tumor synthesis methods.
  \end{abstract}


\end{frontmatter}

\section{Introduction}
%
%
%
%
Cancer is one of the leading causes of deaths worldwide~\cite{ferlay2010estimates}. With the development of computing methods, medical devices and scanners, precision oncology such as early diagnosis and personalized treatment planning can be advanced using diverse medical images~\cite{bi2019artificial}. The training and validation of computerized models for tumor analytics such as automated detection, segmentation, and classification rely on annotated images with domain knowledge. Although a large number of medical images are being collected, acquiring manually annotated tumors is still a time-consuming and labour-intensive task.

Recently, image synthesis techniques for natural image processing achieve significant improvement with deep learning models especially generative adversarial networks (GAN)~\cite{isola2017image}. There are also GAN based methods proposed for the synthesis of lesions in medical images~\cite{jin2018ct,wu2018conditional}. Most of the GANs consist of generative and discriminative networks to capture feature descriptors which represent high-level semantic information of images. The synthetic images are completely new samples, which enhance the feature distribution covered by the original dataset, to increase the data diversity as the synthetic results cover various scales, shapes, and positions~\cite{han2019combining}.

\begin{figure}
  \centering
  \includegraphics[scale=0.55]{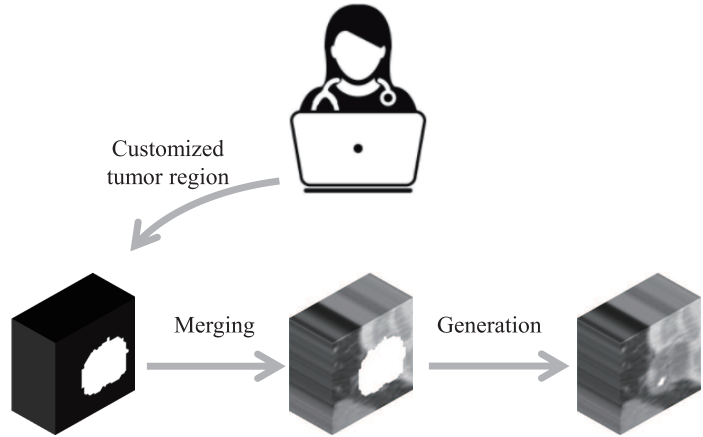}
  \caption{Demonstration of free-form kidney tumor synthesis using the proposed model on a cropped transaxial CT image. The user has the freedom to draw and place a mask of preferred size and shape at an image region. Our model will generate synthetic tumors using user-customized information.}
  \label{fig:overview of our task}
\end{figure}

Most of the current image synthesis models designed for natural image processing, however, cannot satisfy clinical needs. The primary reason is that the types of tumors are diverse which results in various shapes and sizes in different organs such as the liver, lung, and kidney. It is difficult for conventional methods to constrain the appearances and locations of synthesized lesions because images are randomly generated during the creation process. To address this issue, free-form tumor synthesis methods are proposed to enable tumor synthesis by following user-specified requirements including location, shape, and size. The free-form tumor synthesis is a challenging task because the free-form mask (i.e., missing areas) depicted by users could be of arbitrary and irregular shapes (see Figure~\ref{fig:overview of our task}). The second reason is that research for the synthesis of 3-dimensional (3D) tumor volumes is still at the starting stage. That is because 3D tumor synthesis datasets are difficult to obtain and the computational cost for training a 3D model is relatively high. That is why a number of computerized medical image processing models are in 2D instead of explicit 3D. During 2D synthesizing, however, the textural and spatial information may be lost. It could lead inconsistent synthesized results by simply applying a 2D image inpainting model to 3D medical volumes. Thus, there is a great necessity to develop user-customized tumor synthesis models to support the analytics of various lesion of interests and clinical needs.

In this work, we aim to fill in the gap between image synthesis techniques and 3D tumor generation in computed tomography (CT) images, which can provide an alternative solution to the problem of tumor image insufficiency. To address these issues, we propose a free-form tumor/lesion synthesis model using a richer generative adversarial network, named as FRGAN~\footnote{https://github.com/qgking/FRGAN.git, source code will be released publicly.}, to
synthesize tumors using 3D volumetric contextual features. The unique contributions of the proposed model are summarized below.
\begin{itemize}
  \item We formulate the tumor synthesis problem as a free-form inpainting task by using 3D gated convolutions, which allow users such as surgeons or clinicians to generate customized tumors with specific shapes, locations, and sizes. We demonstrate that the synthesized tumors are very similar to real tumors in terms of both qualitative and quantitative validations.
  \item To address the problem of the uncertain boundary between tumor and surrounding tissues during the synthesis process, we propose a new richer convolutional feature enhanced dilated-gated generator. The generator is composed of a dilated-gated encoder to enlarge the perceptive field and a richer convolutional decoder to recover multi-scale features for tumor reconstruction.
  \item To preserve contextual and textural information and enhance the synthetic result, we define a hybrid loss function to penalize the perceptual, style, and multi-mask losses, which contains a wide range of loss functions for medical image synthetic tasks.
\end{itemize}

The paper is structured as follows: Section~\ref{sec:related_works} introduces current synthesis models for medical image applications. We present the proposed FRGAN model in Section~\ref{sec:methodology} and the experimental datasets in Section~\ref{sec:data_set}. The evaluation results and discussions are given in Section~\ref{sec:exp_and_res}.

\section{Related Works}
\label{sec:related_works}
Machine learning techniques are attracting broad research interests and have been proved to be of great potential in advancing computer-aided diagnosis and treatment planning routines~\cite{doi2007computer-aided}. In recent years, we witness the superior performance of machine learning methods such as deep neural networks in various data-driven computer vision applications~\cite{neff2017generative}. To achieve high-quality region of interest (ROI) detection, segmentation, or prediction results, the collection of large numbers of training samples is an essential and significant step. Medical imaging datasets, however, are usually small in data volume, lack of diversity, and harder to acquire when compared with general image datasets. The primary reason is that the annotations of medical images require rigorous domain knowledge from experts~\cite{jin2018ct}. Ethics issues and the protection of patients privacy also pose challenges to the acquisition and publication of patient studies~\cite{neff2017generative}. In order to support effective training and validation of computerized techniques, how to generate new and diverse datasets by fully utilizing the available annotated data is worth investigating.

Recent approaches in solving the scarcity of training images include data augmentation and image synthesis~\cite{zhang2018medical}. Data augmentation has been proved to be a simple and effective way to enlarge training datasets~\cite{yang2018class}. The transformations in data augmentation include rotation, flipping, translation, and elastic distortion~\cite{lafarge2017domain}.
However, conventional transformations cannot significantly increase data diversity. The adversarial-based methods~\cite{mariani2018bagan,luo2018macro,luo2020adversarial} have been proved as an effective approach to address the data scarcity problem, which have achieved huge progress for generating data samples. For instance, Mariani et~al. restored balance in imbalanced datasets via an adversarial balancing GAN (BAGAN) which learned from majority classes and generated images for minority classes~\cite{mariani2018bagan}. Luo et~al. proposed a dual-discriminator based macro-micro adversarial network (MMAN) to improve the human parsing performance~\cite{luo2018macro}. Due to the scarcity of unlabeled target data, Luo et~al. proposed a novel adversarial style mining (ASM) approach for one-shot unsupervised domain adaptation, which proved to be effective in data-scarce scenarios~\cite{luo2020adversarial}. Although the previous approaches cannot be applied directly to the 3D free-form tumor synthesis task, they provide an alternative solution via adversarial learning for image synthesis. Image synthesis models, especially GAN based methods, demonstrated promising results in enhancing data diversity by generating fake realistic images which are similar to real samples~\cite{antoniou2017data,peng2018jointly,yu2017semantic}. GANs have also been applied to medical images such as magnetic resonance imaging (MRI) and CT with different organs including the brain, lung, and liver~\cite{frid2018synthetic,guibas2017synthetic,nie2017medical,wolterink2017deep}.


\textbf{Lung}: In order to generate synthetic lung nodules of different shapes and sizes, Jin et~al.~\cite{jin2018ct} firstly erased image regions containing lung nodules and then passed on the image with erased regions to a cGAN model. The experimental results demonstrated that the synthetic lung nodules can be fused with surrounding lung tissues.
Chuquicusma et~al. generated lung nodules via a deep convolutional GAN (DCGAN) and conducted a visual Turing test for the evaluation of the generated nodules~\cite{chuquicusma2018fool}.

\textbf{Liver}: Ben-Cohen et~al. combined a fully connected neural network (FCN) and a cGAN to generate simulated positron emission tomography (PET) images from given CTs~\cite{ben2019cross}. They proved that the generated PET images could reduce the false-positive rate in liver lesion detection. Ben-Cohen et al. combined the \textit{specific} and \textit{unspecific} representations first before feeding to a generator for new sample synthesis~\cite{ben2018improving}. The experimental results showed an average 7.4\% improvement in terms of liver lesion classification accuracy when the synthetic samples were included for training.

\textbf{Others}: Image synthesis techniques have also been applied to other organs and tasks. For instance, Wolterink et~al. proposed a WGAN model to synthesize blood vessels by parametrizing vessel geometries as 1D signals based on central vessel axis~\cite{wolterink2018blood}. Zijlstra et~al. adopted the cGAN architecture and used the U-Net~\cite{ronneberger2015u} as a generator to synthesize CT images from MRI scans for lower arms in orthopedic applications~\cite{zijlstra2019ct}. Neff et~al. proposed a variant GAN for thorax X-ray image synthesis~\cite{neff2017generative}. The performance of thyroid recognition was improved by using synthetic images when there were insufficient tissue images~\cite{zhang2018medical}. He et~al. improved the performance of conventional GANs in retinal image synthesis from tubular structured annotations by extracting retinal features from multiple layers in VGG~\cite{zhao2018synthesizing}. They also demonstrated the capacity of the proposed model in learning from small datasets.

Even though GAN based approaches and its variants have been used to increase data volume and data variety in different image analysis tasks, the methods mentioned above cannot support user-customized free-form tumor synthesis. Besides, high-quality synthetic results cannot always be  guaranteed because the detailed texture and boundary characteristics may not be preserved during the synthesis process~\cite{abhishek2019mask2lesion,yang2018class}.

To tackle those challenges, we propose a tumor synthesis model with the ability to generate tumors with preserved texture and boundary characteristics, and user-specified sizes, shapes, and locations. We formulate this task as a free-form inpainting problem. Free-form inpainting is attracting research interest in the computer vision community with the recent introduction of 3D gated convolution~\cite{chang2019free,chang2019learnable,yu2018free}. Compared with natural images, a pro-longed challenge in medical image synthesis is that the regions of interests (ROI) such as tumors/lesions are of low resolution and may exhibit indistinct and blurry boundaries.
The novelties of our method are three-fold: (1) To the best of our knowledge, 3D free-form tumor inpainting by dilated and gated convolution is the first of its kind, which supports customized tumor synthesis and serves as a new way for data augmentation; (2) Richer convolutional features are extracted by a newly introduced dilated-gated generator using richer convolutional layers to address the problem of uncertain boundary between tumor and surrounding tissues in the synthesis process; and (3) a hybrid loss function is proposed to preserve texture information while reducing the blurry appearance in order to enhance the authenticity of synthetic tumors.

\begin{figure*}
  \centering
  \includegraphics[scale=0.48]{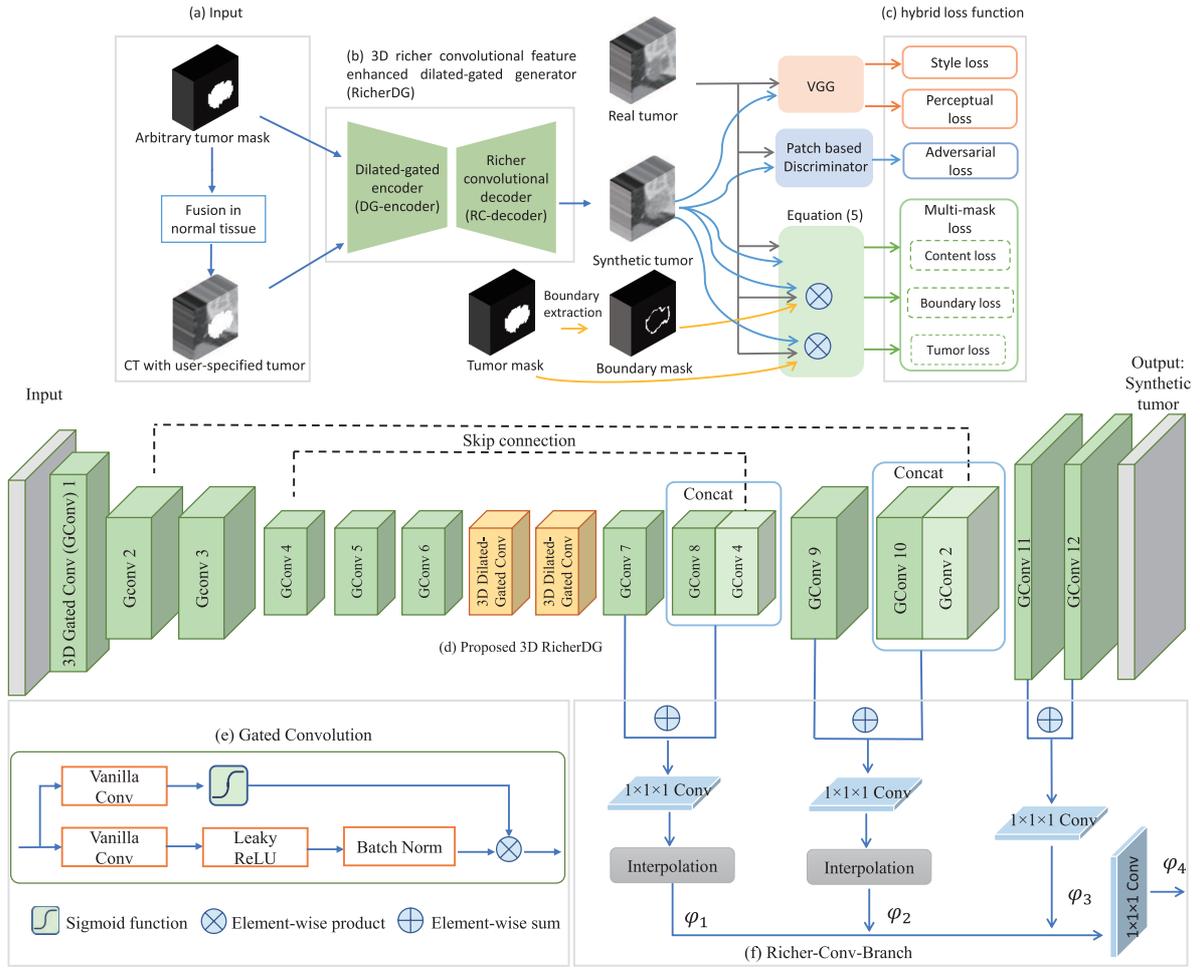}
  \caption{Overview of the proposed FRGAN model. For (a) input mask representing the user-specified tumor to be generated, the mask is firstly fused with the medical image. Then the fused image is sent to the proposed (b) richer convolutional feature enhanced dilated-gated generator (RicherDG) to create the synthetic image. The synthetic tumor is penalized and optimized by (c) the proposed hybrid function in terms of adversarial, multi-mask, perceptual, and style losses. The detailed structure of (b) is given in (d) with the proposed gated convolution operations (GConv) and richer convolutional feature association branch (Richer-Conv-Branch) given in (e) and (f) respectively.}
  \label{fig:overview of architecture}
\end{figure*}

\section{Methods}
\label{sec:methodology}
We formulate the customized synthetic tumor generation task as a free-form inpainting problem. Given an image, the users have the freedom to choose their desired portion of interest (POI). Then the POI is erased by a mask of desired shape and size. Finally, the missing part is restored by a synthetic tumor with a learnt mapping between mask and tumor.

The architecture of the proposed model is given in Figure~\ref{fig:overview of architecture}. The proposed algorithm is based on GAN with three major components including a new dilated-gated generator to enable free-form inpainting, a richer convolutional feature association branch to enhance the authenticity of the synthetic tumor boundary, and a hybrid loss function to incorporate local and global, content and texture information. Given a medical image and a binary mask representing the user-specified shape, location, and size of the tumor, the proposed dilated-gated generator with a richer convolutional feature association branch aims to recover the missing part of the image. Meanwhile, the discriminator evaluates the differences between the real image and recovered synthetic image, which is quantified as an adversarial loss. The training process aims to minimize a new hybrid loss consisting of adversarial loss, multi-mask loss, perceptual loss, and style loss.

\subsection{The basic generative model}
Given an image volume $\boldsymbol{y} \in \mathbb{R}^{X \times Y \times Z}$, where $X, Y$ and $Z$ represent the width ($x$-axis), height ($y$-axis), and depth ($z$-axis) of a patch respectively, the user specification is represented by a binary mask $\boldsymbol{x} \in\left [ 0,1\right ] ^{X \times Y \times Z}$ where lesion is denoted by 1. The mapping between mask and tumor is learnt and formulated based on the concept of a traditional GAN model~\cite{radford2015unsupervised}. GAN consists of a generative network ($G$) and a discriminative network ($D$), where $G: \boldsymbol{x} \in\left [ 0,1\right ] ^{X \times Y \times Z} \rightarrow \hat{\boldsymbol{y}} \in \mathbb{R}^{X \times Y \times Z}$ denotes the tumor synthetic function that takes an erased tumor region of a 3D patch $\boldsymbol{x}$ as input and produces a realistic patch $\hat{\boldsymbol{y}}$ as output. $G$ captures the distributions of the training data and generates the lesions from the input erased masks. In the meanwhile, the performance of $G$ is evaluated by the discriminator $D$, which is designed to classify the synthetic output as real or not. The output of $D$ is considered as a probability map representing the difference between the synthetic distribution and the original distribution. In general, $D$ and $G$ are designed to reach a Nash equilibrium as follows:
\begin{equation}
  \argmin _{G} \argmax _{D} L_{\mathrm{adv}}(G,D).
  \label{eq:gan}
\end{equation}
In the training process, both $G$ and $D$ are trained by following the competing objectives~\cite{goodfellow2014generative}, where the objective of $G$ is to generate ``realistic'' tumors to fool $D$; In the meanwhile, the discriminator $D$ aims to distinguish between the original images and the output of the generator $G$.

\subsection{3D richer convolutional feature enhanced dilated-gated generator (RicherDG)}
To enable customized tumor inpainting, we design a new generator with a dilated-gated encoder (DG-encoder) and a richer convolutional decoder (RC-decoder). The architecture of the new generator is shown in Figure~\ref{fig:overview of architecture}(d) with detailed parameters given in Table~\ref{table:Generator params}. As shown in Figure~\ref{fig:overview of architecture}(d), context information is propagated by the DG-encoder and RC-decoder in a rich long-range skip connection. The DG-encoder extracts pixel-level and object-level features and exploits dilation operations to enlarge the receptive field. The RC-decoder recovers multi-scale features to reconstruct realistic texture appearance within the tumor and near the tumor boundary.

\begin{table*}[]
  \centering
  \caption{Parameter settings of layers in the proposed dilated-gated generator with a richer convolutional feature association branch. Here [ , ] denotes concatenate operation; Conv means convolution; Up stands for up-sampling; GConv denotes gated convolution; DGConv denotes dilated-gated convolution. Note that the Pre-operation column stands for the pre-operation before sending to the corresponding layer in the decoder.}
  \renewcommand\tabcolsep{20.3pt}
  \renewcommand\arraystretch{1.2}
  \begin{tabular}{|c|c|c|c|c|}
    \hline
    \multicolumn{5}{|c|}{Main Part of RicherDG}                                                                                                            \\ \hline
    Encoder               & Output size                            & Decoder                        & Pre-operation     & Output size                      \\ \hline
    Input                 & 2$\times$64\textasciicircum{}3         & DGConv2                        &                   & 256$\times$16\textasciicircum{}3 \\ \hline
    GConv1                & 64$\times$62\textasciicircum{}3        & GConv7                         &                   & 256$\times$16\textasciicircum{}3 \\ \hline
    GConv2                & 128$\times$32\textasciicircum{}3       & GConv8                         &                   & 256$\times$16\textasciicircum{}3 \\ \hline
    GConv3                & 128$\times$32\textasciicircum{}3       & GConv9                         & [GConv8, GConv4]  & 128$\times$32\textasciicircum{}3 \\ \hline
    GConv4                & 256$\times$16\textasciicircum{}3       & GConv10                        &                   & 128$\times$32\textasciicircum{}3 \\ \hline
    GConv5                & 256$\times$16\textasciicircum{}3       & GConv11                        & [GConv10, GConv2] & 64$\times$64\textasciicircum{}3  \\ \hline
    GConv6                & 256$\times$16\textasciicircum{}3       & GConv12                        &                   & 64$\times$64\textasciicircum{}3  \\ \hline
    DGConv1               & 256$\times$16\textasciicircum{}3       & GConv13                        &                   & 1$\times$64\textasciicircum{}3   \\ \hline
    \multicolumn{5}{|c|}{Richer-Conv-Branch}                                                                                                               \\ \hline
    Decoder               & \multicolumn{3}{c|}{ Pre-operation}    & Output size                                                                           \\ \hline
    Conv1                 & \multicolumn{3}{c|}{GConv7+GConv8}     & 1$\times$16\textasciicircum{}3                                                        \\ \hline
    Conv2                 & \multicolumn{3}{c|}{GConv9+GConv10}    & 1$\times$32\textasciicircum{}3                                                        \\ \hline
    Conv3($\varphi _{3}$) & \multicolumn{3}{c|}{GConv11+GConv12}   & 1$\times$64\textasciicircum{}3                                                        \\ \hline
    Up1($\varphi _{1}$)   & \multicolumn{3}{c|}{ Conv1}            & 1$\times$64\textasciicircum{}3                                                        \\ \hline
    Up2($\varphi _{2}$)   & \multicolumn{3}{c|}{ Conv2}            & 1$\times$64\textasciicircum{}3                                                        \\ \hline
    Conv4($\varphi _{4}$) & \multicolumn{3}{c|}{[Conv3, Up1, Up2]} & 1$\times$64\textasciicircum{}3                                                        \\ \hline
  \end{tabular}
  \label{table:Generator params}
\end{table*}

\subsubsection{DG-encoder}
The DG-encoder is composed of 3D gated convolutional layers and 3D dilated-gated convolutional layers.

\textbf{3D gated convolutional layers}: The 3D gated convolutional layers in the encoder replace the commonly used vanilla convolutions which are not fit for the task of free-form inpainting~\cite{yu2018free}. This is because for the normal convolutions in tumor synthesis, invalid voxels belonging to the background region in the binary mask and the valid voxels in the foreground region are stacked together as the input of the neural network. The mixture of features may result in visual artifacts during the testing process, such as color discrepancy, blurriness, and obvious edge responses~\cite{liu2018image}. 3D gated convolution is proposed as shown in Figure~\ref{fig:overview of architecture}(e) to solve this problem. Given input features $I_{X,Y,Z}$, gated convolutional filter $W_{gt}$, and normal convolutional filter $W_{nc}$, the gated convolution can be formulated as:
\begin{equation}
  \begin{split}
    &\mathrm{Gate}_{X,Y,Z}=\sum \sum W_{gt} I_{X,Y,Z},\\
    &\mathrm{Feature}_{X,Y,Z}=\sum \sum W_{nc} I_{X,Y,Z},\\
    &\mathrm{Out}_{X,Y,Z}=\sigma \left ( \mathrm{Gate}_{X,Y,Z}  \right ) f  \left ( \mathrm{Feature}_{X,Y,Z} \right ),
  \end{split}
  \label{eq:gate mechanism}
\end{equation}
where $\sigma$ is the Sigmoid function to normalize the gate value between 0 and 1, and $f$ denotes the activation function.

\textbf{3D dilated-gated convolutional layers}: The layers with dilated convolutions are used to broaden the receptive field. There are 2 layers that apply $3 \times 3 \times 3$ convolutions with dilation factors at 2 and 4.
The kernel size increases by increasing this dilation factor~\cite{yu2015multi}. Therefore, the receptive field can be effectively enlarged through layers and more information can be extracted by dilated convolutions.


\subsubsection{RC-decoder}
As shown in Figure~\ref{fig:overview of architecture}(d), the RC-decoder is composed of 3D dilated-gated convolutions, 3D gated convolutions, and a newly proposed richer convolutional feature association branch (Richer-Conv-Branch). The Richer-Conv-Branch is proposed to enlarge the perception fields and recover pixel-wise synthesis from low-level and object-level feature maps. This is motivated by previous research that multi-scale information is useful for edge detection with the assistance of different sizes of receptive fields in the backbone branch~\cite{liu2017richer}.

Each set of 3D gated convolution is composed of two gated convolutional layers. The Richer-Conv-Branch consists of a $1 \times 1 \times 1$ vanilla convolutional layer and an interpolation layer to accumulate the outputs of gated convolutions and associated multi-level boundary features. For the first two convolutional sets in the decoding process, the output of each 3D gated convolutional layer is concatenated to a $1 \times 1 \times 1$ vanilla convolutional layer. Then, trilinear interpolation is performed to up-sample the feature maps. For the last convolution in the decoder, the gated convolutional layer is first connected to a $1 \times 1 \times 1$ convolutional layer. Finally, the outputs of the three sets of convolutions are concatenated together and fed into a $1 \times 1 \times 1$ convolutional layer to fuse the feature maps.

\subsection{Patch-based discriminator}
We exploit patch-based discriminator~\cite{isola2017image} as $D$ in our model to capture multi-scale global and local features for blur reduction. Different from the commonly used discriminator in GAN~\cite{han2018gan}, patch-based discriminator randomly selects $N \times N \times N$ patches from the input and classifies the selected patches as `real' or `fake'. By doing so, the discriminator is restricted to focus on the high-frequent structures because it only penalizes features at the range of patches.

The discriminator is composed of four blocks and there are a convolutional layer, a Leaky ReLU, and batch normalization in each block. The output of the discriminator represents the probability of whether the input data is derived from the distribution of real tumors/lesions. The output of the discriminator is named as adversarial loss ($L_{\mathrm{adv}}$) as defined in Equation~(\ref{eq:adversarial}):
\begin{equation}
  \begin{split}
    L_{\mathrm{adv}}=\argmin _{G} \argmax _{D} (&\mathbb{E}_{\boldsymbol{x},\boldsymbol{y}}[\log D(\boldsymbol{x},\boldsymbol{y})]\!+ \\
    &\mathbb{E}_{\boldsymbol{x}}[\log (1-D(\boldsymbol{x},G(\boldsymbol{x})))]).
  \end{split}
  \label{eq:adversarial}
\end{equation}
In our work, we use the binary cross entropy loss as the objective function to classify the input tumor as real or fake.

\subsection{Hybrid loss function}
Apart from the adversarial loss, we propose the multi-mask loss ($L_{\mathrm{mm}}$), and introduce the style loss ($L_{\mathrm{sty}}$) and perceptual loss ($L_{\mathrm{percep}}$) to capture the distribution of the erased tumor/lesion area (i.e., boundary and tumor). A hybrid loss function is defined to associate the losses as
\begin{equation}
  L_{\mathrm{GAN}}=L_{\mathrm{adv}}+\lambda L_{\mathrm{mm}}+\delta L_{\mathrm{percep}} +\eta L_{\mathrm{sty}},
  \label{eq:final loss formulation}
\end{equation}
where $\lambda$, $\delta$, and $\eta$ are weights to balance different losses.

\subsubsection{Multi-mask loss}

$L_{\mathrm{mm}}$ is composed of content-wise loss ($L_{\mathrm{cw}}$), synthetic tumor loss ($L_{\mathrm{st}}$), and  boundary loss ($L_{\mathrm{sb}}$) as defined in Equation~(\ref{eq:multi-mask loss formulation}):
\begin{equation}
  L_{\mathrm{mm}}=\alpha L_{\mathrm{cw}} + \beta L_{\mathrm{st}} + \gamma L_{\mathrm{sb}},
  \label{eq:multi-mask loss formulation}
\end{equation}
where $\alpha, \beta$, and $\gamma$ are the weight scalars, $L_{\mathrm{cw}}$, $L_{\mathrm{st}}$, $L_{\mathrm{sb}}$ are defined to represent the overall appearance of a patch, the features within a synthetic tumor, and boundary between tumor and surrounding tissues, respectively.

Given the input tumor mask, let $M_{\mathrm{st}}$ be the binary mask where the voxels within the erased tumor volume are filled by 1, $M_{\mathrm{sb}}$, which is obtained by inverted binary erosion with Gaussian filtering operations, represents the surrounding boundary region of $M_{\mathrm{st}}$, and the content-wise loss $L_{\mathrm{cw}}$ is calculated as $\mathbb{E}_{\boldsymbol{x},\boldsymbol{y}}\left[|y-G(\boldsymbol{x})|\right]$. $L_{\mathrm{st}}$ is measured as $\mathbb{E}_{\boldsymbol{x}, \boldsymbol{y}}\left[M_{\mathrm{st}} |\boldsymbol{y}-G(\boldsymbol{x})|\right]$. For the calculation of $L_{\mathrm{sb}}$, as the Richer-Conv-Branch in RC-decoder can extract richer multi-scale convolutional features~\cite{liu2017richer} near the tumor boundary, $L_{\mathrm{sb}}$  is defined as,
\begin{equation}
  L_{\mathrm{sb}}= \sum_{s=1}^{S} \mathbb{E}_{\boldsymbol{x}, \boldsymbol{y}}\left[M_{\mathrm{sb}} |\boldsymbol{y}-\varphi_{s}|\right],
  \label{eq:final L_sb formulation}
\end{equation}
where $\varphi_{s}$ denotes the $s$'s output of the decoder as shown in Figure~\ref{fig:overview of architecture}(f). $s$ is in $\{ 1,...,S \}$ and $S$ stands for the side output number of Richer-Conv-Branch. We set $S$ as 4 in our network. To be more specific, $\varphi_{s}$ comes from Up1, Up2, Conv3, and Conv4 in Table~\ref{table:Generator params}. An additional fusion feature map is concatenated to enhance the edge map of objects according to~\cite{liu2017richer}.

\subsubsection{Perceptual loss}
As the multi-mask loss function may fail to preserve the perceptual quality, a perceptual loss is introduced~\cite{gatys2015neural}. In addition, the perceptual loss is designed to alleviate the blurriness that may appear around the tumor boundary during the reconstruction process~\cite{liu2018image,pathak2016context,nazeri2019edgeconnect}. The perceptual loss is calculated by minimizing the following norm of the difference between the generated tumor and the real tumor:
\begin{equation}
  L_{\mathrm{percep}}=\!\sum_{d \in \{ 1,...,Z \}, p \in \{ 1,...,P\}} \!\frac{1}{X_{d} Y_{d}}\left\|\phi^{\left(p\right)}(\boldsymbol{y}_{d})-\phi^{\left(p\right)}(\hat{\boldsymbol{y}}_{d})\right\|_{1},
  \label{eq:perceptual loss}
\end{equation}
where $\phi^{\left(p\right)}$ denotes the activation from the $p$th layers of the pre-trained network (i.e., VGG in our method), $P$ is the total number of layers of the pre-trained network. We use layers $\mathrm{ReLU}_{2\_2}$, $\mathrm{ReLU}_{3\_3}$, and $\mathrm{ReLU}_{4\_3}$ to calculate the perceptual loss.

\subsubsection{Style loss}
Style loss~\cite{gatys2015neural} is introduced to minimize the deviation between the target $\boldsymbol{y}$ and generated $\hat{\boldsymbol{y}}$ to transfer the style from $\boldsymbol{y}$ to $\hat{\boldsymbol{y}}$. The style loss, which is similar to perceptual loss, is obtained as the difference between the Gram matrices of two objects:
\begin{equation}
  \begin{split}
    &\mathbb{G}_{d}^{\left(p\right)}\left(\boldsymbol{y}_{d}\right)=\left(\phi^{\left(p\right)}(\boldsymbol{y}_{d})\right)^{T} \left(\phi^{\left(p\right)}(\boldsymbol{y}_{d})\right) \\
    &L_{\mathrm{sty}}\!=\!\sum_{d \in \{ 1,...,Z \}, p \in \{ 1,...,P \}} \!\frac{1}{ p^{2} X_{d} Y_{d}}\left\|\mathbb{G}_{d}^{\left(p\right)}\left(\boldsymbol{y}_{d}\right)- \mathbb{G}_{d}^{\left(p\right)}\left(\hat{\boldsymbol{y}}_{d}\right)\right\|_{1},
  \end{split}
  \label{eq:style loss}
\end{equation}
where $\mathbb{G}_{d}^{\left(p\right)}\left(\boldsymbol{y}_{d}\right)$ denotes the Gram matrix of the output of the $p$th layer with the $d$th input.

\subsection{Qualitative and quantitative evaluation methods}
The effectiveness of the proposed tumor synthesis model in medical image analysis is qualitatively and quantitatively evaluated. Ablation study is also performed to investigate the contributions of the major components in our model. As mentioned in~\cite{yu2018free}, the lack of quantitative evaluation metrics for the evaluation of synthesis models is a common issue in natural images, which is also a challenging task in medical images. Thus, we perform an indirect evaluation of synthetic models via other image processing tasks. In this work, we evaluate the synthesizing performance by a segmentation task similar to~\cite{zhao2018synthesizing}.

Our hypothesis is that if the segmentation accuracy of a segmentation model, for instance, U-Net~\cite{ronneberger2015u,cciccek20163d}/Attention U-Net~\cite{oktay2018attention} (AU-Net)/Med3D~\cite{chen2019med3d}/U2-Net~\cite{qin2020u2}, trained on a dataset composed of both real and synthetic images is better than that on a dataset of purely real images, it can be inferred that synthetic images contribute to solving the problem of data insufficiency. When comparing two different synthesis models, if one U-Net trained using real images and synthetic images generated by model A outperforms another U-Net trained using the same real images and synthetic images generated by model B, we draw the conclusion that model A is more effective than model B for image synthesis.

Given an image dataset, for all the cases we divide the dataset into training (60\% real tumors), validation (20\% real tumors), and testing (20\% real tumors) subsets. For each synthesis model, the segmentation model is trained using a set of training data (60\% real tumors) and synthetic data generated by the model using validation data (20\% real tumors). The segmentation accuracy is evaluated over the testing data (20\% real tumors). The evaluation metrics include Dice similarity score (DSC), Jaccard similarity coefficient (Jaccard), volume overlap error (VOE), relative volume difference (RVD), and Hausdorff distance (HD)~\cite{HausdorffDistance,bilic2019liver}.

\section{Dataset}
\label{sec:data_set}
\subsection{Dataset}
We perform a comprehensive evaluation of our model using three public CT datasets covering liver, kidney tumors, and lung nodules.
\subsubsection{Liver Tumor Segmentation Challenge dataset}
The Liver Tumor Segmentation Challenge (LiTS)~\cite{bilic2019liver} provides 130 CT scans with ground truth for training and 70 scans without ground truth for testing which are collected from different hospitals. All the CT images are with 512$\times$512 in-plane resolution, while slice thickness across patient scans varies. There are 116 scans containing tumors among all the 130 CT scans. We used the 116 training scans with corresponding manual tumor delineations for our study.


\subsubsection{Kidney Tumor Segmentation Challenge dataset}
The Kidney Tumor Segmentation Challenge (KiTS) dataset~\cite{heller2019kits19} contains 300 CT scans where 210 scans are with manual kidney tumor segmentations. All the CT images are of 512$\times$512 in-plane resolution. In this work, we used 210 scans for model training and testing.

\subsubsection{Lung Nodule Analysis Challenge dataset}
The Lung Nodule Analysis 2016 (LUNA) Challenge dataset~\cite{setio2017validation} contains annotated data for 888 patients. For each patient, there are CT images and an XML file with nodule locations on each 512 $\times$ 512 slice. The nodule locations are a series of coordinates which contain the center of the lesion and the diameter of the lesion. As the annotations of lesions are in coarse format which is different from the other two datasets, the generation of user-specified masks are different from the others.  We will describe the generation of masks in detail together with experimental results in Section~\ref{sec:exp_and_res}.

\subsection{Data pre-processing and 3D tumor cube generation}
The Hounsfield units (HU) of raw CT images in KiTS, LiTS, and LUNA are in the range of $\left [-1000, 1000 \right ]$. To coarsely remove irrelevant objects, we firstly perform a global cut-off using HU windows at $\left [-200, 300 \right ]$, $\left [-100, 200 \right ]$, and $\left [-1000, 600 \right ]$ for KiTS, LiTS, and LUNA respectively. Then we normalize each slice to zero mean~\cite{kamnitsas2017efficient} and linearly scale the processed intensity values in all images to $\left [0,1 \right ]$. Furthermore, 3D connected component labeling~\cite{hossam20103d} is performed for each case to remove small tumors with very limited contextual information. We empirically define that the tumors containing less than 200, 400, and 100 pixels are the small ones to be filtered out on KiTS, LiTS, and LUNA respectively.

After the above pre-processing, a 3D tumor cube $\boldsymbol{y}$ is generated by cropping a cube centered on each tumor with a padding of 20 pixels in all three dimensions. The input patch $\boldsymbol{x}$ is acquired by erasing the tumor pixels from $\boldsymbol{y}$. All patches are then resampled to a fixed size of $64 \times 64 \times 64$ and randomly flipped and rotated by ($0^{\circ}$, $90^{\circ}$, $180^{\circ}$, and $270^{\circ}$) in three-orthogonal directions. The datasets are divided into a 60\% training, 20\% validation, and 20\% testing sets. In total, there are 1,356/2,344/2,748 tumor samples which are extracted from KiTS/LiTS/LUNA, where the training dataset consists of 792/1,580/1,650 samples from the first  126/69/315 scans, the validation dataset consists of 258/168/549 samples from the following 42/23/105 scans, and the test set consists of 306/596/549 patches from the remaining 42/24/106 scans.

\begin{figure}
  \centering
  \includegraphics[scale=1.3]{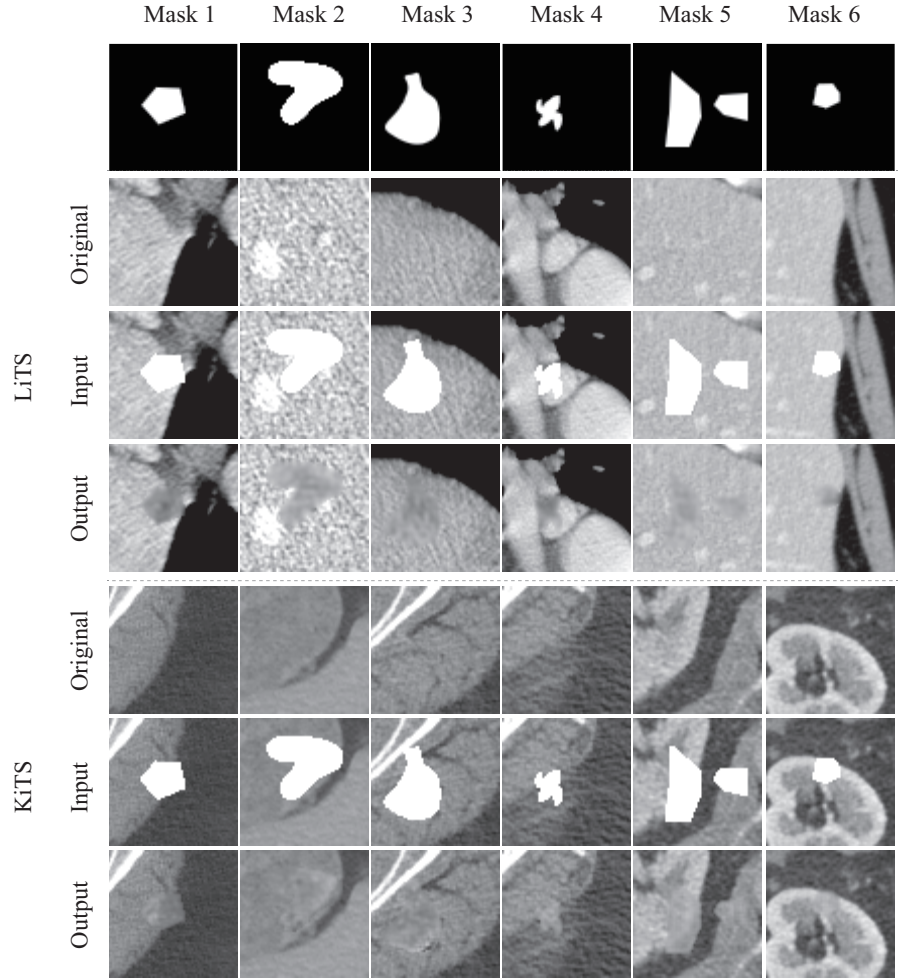}
  \caption{Synthesis results using six different input masks on liver and kidney CT images from LiTS and KiTS.}
  \label{fig:synthe_mask}
\end{figure}

\begin{figure}
  \centering
  \includegraphics[scale=1.2]{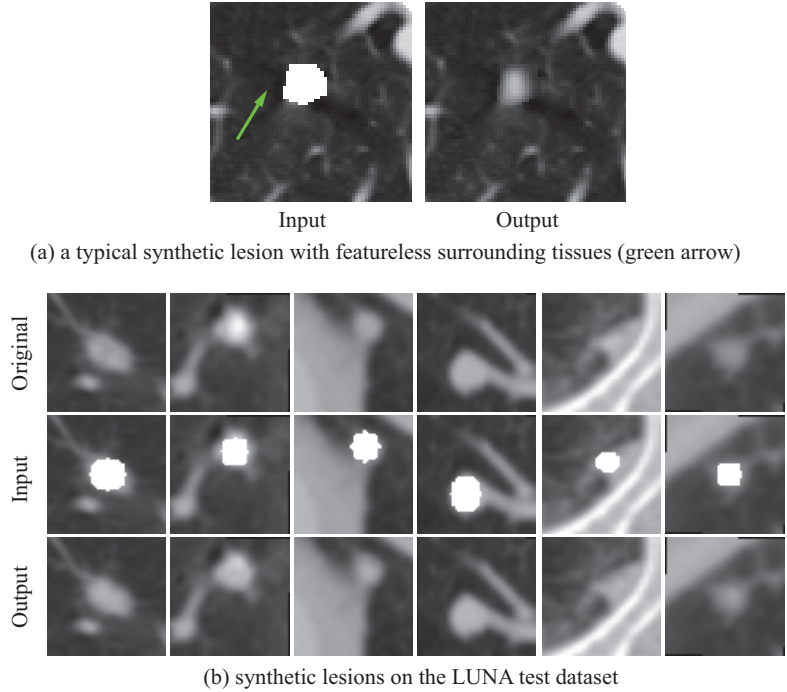}
  \caption{Synthesis results of lung nodule CT images from the LUNA test dataset. (a) one case when there are non-informative surrounding tissues. (b) six cases using different masks on the test dataset.}
  \label{fig:synthe_luna}
\end{figure}

\section{Experiments}
\label{sec:exp_and_res}
\subsection{Parameter settings}
The FRGAN architecture was implemented by the PyTorch~\cite{paszke2017automatic} library and was trained on four NVIDIA 1080Ti GPUs. All the models were trained from scratch. The loss function was optimized by the Adam optimizer. The initial learning rate was 0.0001. For the loss function, $\alpha, \beta$, $\gamma$, $\lambda$, $\delta$, and $\eta$ were set as 1, 10, 1, 1, 1, and 100 respectively. The $\delta$ and $\eta$ were set the same as~\cite{chang2019free}, the other parameters were set mainly for balancing the weight of different losses during the training stage. For all the synthesis models in comparison, 100 epochs were performed to train the models. The Adam solver with a batch size of 4 was applied to minimize the objectives. During training, $\gamma$ linearly increased from 0 to 1 in the first 30 epochs and then remained at 1 for the remaining epochs.

\subsection{Free-form synthesized tumors by FRGAN}
The examples of synthetic tumors generated by our model using liver and kidney CT images from LiTS and KiTS with six different masks are given in Figure~\ref{fig:synthe_mask}. Tumors can be synthesized in arbitrary shapes in order to increase the diversity of data using the proposed method. The synthetic results retain consistent textural and color features. It is shown that the synthetic tumors were fused to the CT images naturally.

For the lung nodule CT images, the annotations of lesions are a series of lesion center coordinates and diameters. Thus, the accurate voxels within a lesion cannot be fully extracted. In addition, when the lung nodules are isolated in the lung region as shown in Figure~\ref{fig:synthe_luna}(a), if we generated a mask by giving the center coordinate and diameter, the dark non-nodule region will be included as foreground. As a result, we found that the generated synthetic lesions were smaller than expected as shown by the output image in Figure~\ref{fig:synthe_luna}(a) because the dark background region was labeled as desired foreground. Therefore, we choosed 6 masks with different sizes and shapes as shown in Figure~\ref{fig:synthe_luna}(b) on the LUNA test dataset. The experimental results demonstrated that when there were informative surrounding tissues near the mask, our model was able to generate synthetic lesions effectively for the images from LUNA.

In summary, the qualitative results show that the FRGAN can generate realistic synthetic tumors in various appearances, textures, and locations, and even when the input mask is irregular and complicated.

\subsection{Quantitative analysis}
\subsubsection{Comparison with other methods}
To evaluate the contribution of synthetic images in tumor segmentation, we firstly train baseline U-Net~\cite{ronneberger2015u}, AU-Net~\cite{oktay2018attention}, Med3D~\cite{chen2019med3d}, and U2-Net~\cite{qin2020u2} models. For fair comparisons, all the models are trained using 50 epochs with the Adam optimizer. The optimizer is decayed by a polynomial learning rate policy, where the initial learning rate (0.0001) is multiplied by $\left(1-\frac{iter}{total\_{iter}}\right)^{0.9}$. The weights after the model converges are used for testing. As in all the cases, a dataset is split into training, validation, and testing sets using 60\%, 20\%, and 20\% of real data, the baseline models are trained using the training set (60\% real data). The segmentation performance over the testing dataset (20\% real data) is given in Table~\ref{table:segmentation_performance}. Each of the baseline models obtained the worst spatial overlap measurements over LUNA, where the Dice, Jaccard, and VOE are 0.711/0.713/0.696/0.711, 0.573/0.575/0.558/0.572, and 0.428/0.425/0.442/0.428, respectively. The best spatial overlap was achieved in KiTS with Dice, Jaccard, and VOE of 0.840/0.864/0.818/0.847, 0.737/0.771/0.705/0.748, and 0.263/0.229/0.295/0.252.

The segmentation results using both real data and synthetic images are obtained by training a U-Net~\cite{ronneberger2015u}/AU-Net~\cite{oktay2018attention}/Med3D~\cite{chen2019med3d}/U2-Net~\cite{qin2020u2} using the same 60\% real data and synthetic images generated by different synthesis models using the validation set. The proposed model is compared with two state-of-the-art models. The model in~\cite{jin2018ct} is a cGAN model to generate 3D lung nodules of different sizes and the model in~\cite{abhishek2019mask2lesion} is based on the pix2pix model~\cite{isola2017image} with the original purpose to generate masked skin lesions.

\begin{landscape}
  \begin{table*}[]
    \centering
    \caption{Quantitative evaluation of the synthesizing results from liver CT, lung nodule CT, and kidney CT images using different synthesis methods. Real refers to training a U-Net/AU-Net/Med3D/U2-Net using the training dataset composed of 60\% real tumor. Real + synthesis model indicates the training of a U-Net/AU-Net/Med3D/U2-Net using real tumor and synthetic images generated by the synthesis model using validation dataset (20\% real tumor).}
    \label{table:segmentation_performance}
    \renewcommand\arraystretch{1.7}
    \renewcommand\tabcolsep{4.0pt}
    \begin{tabular}{ccccccc|ccccc|ccccc}
      \toprule
      \multirow{2}{*}{Dataset}                          & \multirow{2}{*}{Training data}       & \multicolumn{5}{c|}{LiTS} & \multicolumn{5}{c|}{LUNA} & \multicolumn{5}{c}{KiTS}                                                                                                       \\ \cline{3-17}
                                                        &                                      & Dice                      & Jaccard                   & VOE                      & RVD   & HD    & Dice  & Jaccard & VOE   & RVD    & HD    & Dice  & Jaccard & VOE   & RVD    & HD    \\ \hline
      \multirow{4}{*}{U-Net~\cite{ronneberger2015u}}    & Real                                 & 0.725                     & 0.594                     & 0.406                    & 0.468 & 1.899 & 0.711 & 0.573   & 0.427 & 0.337  & 1.458 & 0.840 & 0.737   & 0.263 & -0.028 & 2.661 \\
                                                        & Real+~\cite{abhishek2019mask2lesion} & 0.728                     & 0.597                     & {\color{red}{\textbf{0.403}}}                    & 0.692 & {\color{red}{\textbf{1.886}}} & 0.716 & 0.577   & 0.423 & 0.548  & 1.451 & 0.855 & 0.759   & 0.241 & -0.011 & 2.555 \\
                                                        & Real+~\cite{jin2018ct}               & 0.726                     & 0.595                     & 0.405                    & 0.495 & 1.891 & {\color{red}{\textbf{0.719}}} & {\color{red}{\textbf{0.580}}}   & {\color{red}{\textbf{0.420}}} & 1.247  & 1.448 & 0.853 & 0.756   & 0.244 & -0.019 & 2.570 \\
                                                        & Real+FRGAN                           & {\color{red}{\textbf{0.730}}}                     & {\color{red}{\textbf{0.599}}}                     & 0.404                    & {\color{red}{\textbf{0.398}}} & 1.895 & 0.716 & 0.577   & 0.423 & {\color{red}{\textbf{0.334}}}  & {\color{red}{\textbf{1.442}}} & {\color{red}{\textbf{0.858}}} & {\color{red}{\textbf{0.763}}}   & {\color{red}{\textbf{0.237}}} & {\color{red}{\textbf{-0.003}}} & {\color{red}{\textbf{2.529}}} \\ \hline

      \multirow{4}{*}{AU-Net~\cite{oktay2018attention}} & Real                                 & 0.717                     & 0.588                     & 0.412                    & 3.935 & 1.888 & 0.713 & 0.575   & 0.425 & {\color{red}{\textbf{0.446}}}  & 1.445 & 0.864 & 0.771   & 0.229 & -0.009 & 2.521
      \\
                                                        & Real+~\cite{abhishek2019mask2lesion} & 0.720                     & 0.591                     & 0.409                    & 0.744 & 1.877 & 0.712 & 0.574   & 0.426 & 2.336  & 1.446 & 0.865 & 0.773   & 0.227 & {\color{red}{\textbf{-0.002}}} & 2.520
      \\
                                                        & Real+~\cite{jin2018ct}               & 0.723                     & 0.594                     & 0.406                    & 0.749 & 1.882 &0.712 & 	0.573 &	0.427 &	0.675& 	1.448   & 0.862 & 0.770   & 0.230 & 0.004  & 2.522
      \\
                                                        & Real+FRGAN                           & {\color{red}{\textbf{0.725}}}                     & {\color{red}{\textbf{0.596}}}                     & {\color{red}{\textbf{0.404}}}                    & {\color{red}{\textbf{0.542}}} & {\color{red}{\textbf{1.876}}} & {\color{red}{\textbf{0.716}}} & {\color{red}{\textbf{0.577}}}   & {\color{red}{\textbf{0.423}}} & 0.447  & {\color{red}{\textbf{1.435}}} & {\color{red}{\textbf{0.869}}} & {\color{red}{\textbf{0.779}}}   & {\color{red}{\textbf{0.221}}} & -0.011 & {\color{red}{\textbf{2.508}}}
      \\ \hline
      \multirow{4}{*}{Med3D~\cite{chen2019med3d}}       & Real                                 & 0.710                     & 0.573                     & 0.427                    & 0.410 & 2.013 & 0.696 & 0.558   & 0.442 & 0.370  & 1.551 & 0.818 & 0.705   & 0.295 & -0.043 & 2.802
      \\
                                                        & Real+~\cite{abhishek2019mask2lesion} & 0.711                     & 0.574                     & 0.426                    & 0.362 & {\color{red}{\textbf{2.004}}} & {\color{red}{\textbf{0.705}}} & {\color{red}{\textbf{0.567}}}   & {\color{red}{\textbf{0.433}}} & {\color{red}{\textbf{0.350}}}  & 1.533 & 0.825 & 0.715   & 0.285 & -0.033 & 2.760
      \\
                                                        & Real+~\cite{jin2018ct}               & 0.712                     & 0.575                     & 0.425                    & 0.350 & 2.005 & 0.703 & 0.566   & 0.434 & 0.479  & 1.533 & {\color{red}{\textbf{0.827}}} & {\color{red}{\textbf{0.717}}}   & {\color{red}{\textbf{0.283}}} & {\color{red}{\textbf{-0.030}}} & {\color{red}{\textbf{2.751}}}
      \\
                                                        & Real+FRGAN                           & {\color{red}{\textbf{0.714}}}                     & {\color{red}{\textbf{0.577}}}                     & {\color{red}{\textbf{0.423}}}                    & {\color{red}{\textbf{0.309}}} & 2.014 & 0.703 & 0.566   & 0.434 & 0.521  & {\color{red}{\textbf{1.528}}} & {\color{red}{\textbf{0.827}}} & {\color{red}{\textbf{0.717}}}   & {\color{red}{\textbf{0.283}}} & -0.034 & 2.753
      \\ \hline
      \multirow{4}{*}{U2-Net~\cite{qin2020u2}}          & Real                                 & 0.736                     & 0.605                     & 0.395                    & 0.285 & 1.885 & 0.711 & 0.572   & 0.428 & 0.178  & 1.501 & 0.847 & 0.748   & 0.252 & -0.015 & 2.595
      \\
                                                        & Real+~\cite{abhishek2019mask2lesion} & 0.733                     & 0.603                     & 0.397                    & 0.401 & 1.862 & 0.716 & 0.576   & 0.424 & 0.194  & 1.479 & 0.851 & 0.755   & 0.245 & -0.014 & 2.556
      \\
                                                        & Real+~\cite{jin2018ct}               & 0.737                     & 0.607                     & 0.393                    & {\color{red}{\textbf{0.284}}} & 1.880 & 0.712 & 0.572   & 0.428 & 0.953  & 1.483 & 0.850 & 0.752   & 0.248 & {\color{red}{\textbf{-0.005}}} & 2.564
      \\
                                                        & Real+FRGAN                           & {\color{red}{\textbf{0.748}}}                     & {\color{red}{\textbf{0.618}}}                     & {\color{red}{\textbf{0.382}}}                    & 1.310 & {\color{red}{\textbf{1.852}}} & {\color{red}{\textbf{0.719}}} & {\color{red}{\textbf{0.579}}}   & {\color{red}{\textbf{0.422}}} & {\color{red}{\textbf{0.160}}}  & {\color{red}{\textbf{1.470}}} & {\color{red}{\textbf{0.858}}} & {\color{red}{\textbf{0.762}}}   & {\color{red}{\textbf{0.238}}} & -0.009 & {\color{red}{\textbf{2.519}}}
      \\
      \bottomrule
    \end{tabular}
    \begin{tablenotes}[flushleft]
      \item {\color{red}{\textbf{Red: Best response}}}
      \end{tablenotes}
  \end{table*}
\end{landscape}

As shown in Table~\ref{table:segmentation_performance}, when using extra synthetic images generated by synthesis models, the overall segmentation performance was improved in terms of Dice, Jaccard, VOE, RVD, and HD over all the three datasets. Specifically, the AU-Net/U2-Net trained with tumors generated by our method achieved the best performance over LiTS and KiTS datasets in terms of spatial overlap measured by Dice, Jaccard, VOE, and HD when compared to those trained with tumors generated by ~\cite{jin2018ct} and~\cite{abhishek2019mask2lesion}. For LiTS, the advanced U2-Net obtained the best performance with our synthetic samples, which outperformed the second best by 1.1\%/1.1\% in terms of Dice/Jaccard. For LUNA, the models augmented with synthetic tumors outperformed the corresponding baseline models without synthetic tumors. The most significant improvement, which improved the Dice/Jaccard by 1.8\%/2.6\% and lowered the VOE by 2.6\%, was achieved by U-Net on the KiTS dataset.

Overall, U2-Net achieved competitive performance among all the models with those tumors synthesized by FRGAN. The quantitative evaluations prove the capacity of our proposed FRGAN on synthesizing critical pathological information, which augmented the segmentation performance.


\begin{table*}[t]
  \centering
  \caption{Jaccard performance achieved by U-Net with different proportions of the synthetic samples in the validation data (20\% of total dataset). $M$\%Syn+$N$\%Real indicates the $M$\% of the validation data is the synthetic data generated by our FRGAN, and the remaining $N$\% of the validation data is the real tumors. Note that 0\%Syn+0\%Real contains only 60\% real data for training.}
  \label{table:different_syn_percentage}
  \renewcommand\arraystretch{1.2}
  \renewcommand\tabcolsep{10.8pt}
  \begin{tabular}{ccccccc}
    \toprule
    Dataset & \tabincell{c}{0\%Syn+                                         \\0\%Real}                          & \tabincell{c}{0\%Syn+ \\100\%Real}  & \tabincell{c}{25\%Syn+ \\75\%Real}    & \tabincell{c}{50\%Syn+ \\50\%Real}    & \tabincell{c}{75\%Syn+ \\25\%Real}  & \tabincell{c}{100\%Syn+ \\0\%Real}    \\ \hline
    LiTS    & 0.594                 & 0.599 & 0.596 & 0.597 & 0.598 & 0.599 \\ \hline
    LUNA    & 0.573                 & 0.586 & 0.584 & 0.579 & 0.581 & 0.577 \\ \hline
    KiTS    & 0.737                 & 0.763 & 0.758 & 0.761 & 0.759 & 0.763 \\

    \bottomrule
  \end{tabular}
\end{table*}

\subsubsection{Investigation of different amounts of synthetic data}
We further investigate how much synthetic data are needed to obtain satisfactory results by changing different amounts of synthetic data. We compared segmentation performance of U-Net on 60\% real data and 20\% validation data, which included different amounts of synthetic data and real data. The Jaccard score evaluations are shown in Table~\ref{table:different_syn_percentage}. As shown, the baseline model, trained on 60\% real data only, achieved the lowest Jaccard performance when compared to those trained with extra 20\% validation data. When augmented with 20\% real validation data, the performance gained on Jaccard was consistent with 0.5\%, 1.3\%, and 2.6\% improvement on LiTS, LUNA, and KiTS, respectively.
As we gradually changed the different proportions of synthetic data, the performance remained stable with minor perturbations on the LiTS and KiTS datasets. The stability indicated that the synthetic data samples contained pathological information as the real ones. We also find that the performance on LUNA dropped continuously as the proportion of synthetic data increased when compared to the 100\% real data. As aforementioned, the annotations of lesions are a series of lesion center coordinates and diameters, the synthetic region can have non-lesion regions, which brings artifacts in accurate lesion segmentation. The larger the proportion of the synthetic lesion, the higher the probability that the performance may drop.

In summary, the synthetic data provides a form of augmentation, allowing for much less ``real'' data to be used to achieve a comparable level of performance for clinical applications.

\subsection{Qualitative analysis}
\begin{figure*}
  \centering
  \includegraphics[scale=0.8]{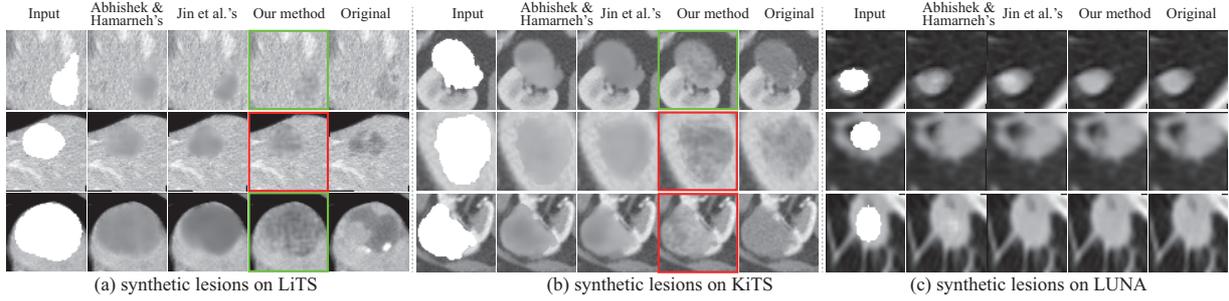}
  \caption{Synthesis results of three different datasets covering liver, kidney, and lung, and twelve different shapes of masks using our model and two other methods~\cite{abhishek2019mask2lesion,jin2018ct} in comparison. The FRGAN synthesizes tumors with less blurriness than other methods on the test dataset. Cases highlighted by green boxes are cases with diverse tumor contents. Tumors in red boxes are those with richer boundary information.}
  \label{fig:synthe_comp}
\end{figure*}

Examples of the synthetic images generated by our model and two other models in comparison are given in Figure~\ref{fig:synthe_comp}. As shown, the synthetic tumors and lesions generated by the proposed FRGAN are of better quality and full of patterns that may occur in real tumors. In comparison, the synthetic tumors generated by the other two methods are observed to be blurry and smooth in appearance without rich texture features. For the synthesis results on the LUNA dataset, it is not obvious to observe significant visual differences even though the quantitative evaluation results in Table~\ref{table:segmentation_performance} are statistically obvious.

Even though the model in~\cite{abhishek2019mask2lesion} and the model in~\cite{jin2018ct} can also support user-customized tumor synthesis, our model improved the visual appearance of the synthesized tumors using the newly introduced RicehrGonv-Branch and hybrid loss functions as shown in Figure~\ref{fig:synthe_comp}. The original model in~\cite{abhishek2019mask2lesion} is designed for 2D skin lesions from dermoscopic images. The boundary between lesion and surrounding tissues on dermoscopic images is discernible when compared with the soft tissue boundaries on grayscale CT. Thus, it can be seen from Figure~\ref{fig:synthe_comp} that the 3D tumors generated by the method in~\cite{abhishek2019mask2lesion} are blurry when compared with our proposed FRGAN. The tumors generated by FRGAN are natural with high visual texture similarities with surroundings. When compared with~\cite{jin2018ct}, the FRGAN introduces RicherConv-Branch to enhance the synthesis of tumor boundaries. In addition, the hybrid loss function further reinforced tumor synthesis as demonstrated by the improved tumor quality in Figure~\ref{fig:synthe_comp} that our model restored richer contextual information when compared with the other two models.

\begin{figure*}
  \centering
  \includegraphics[scale=1.6]{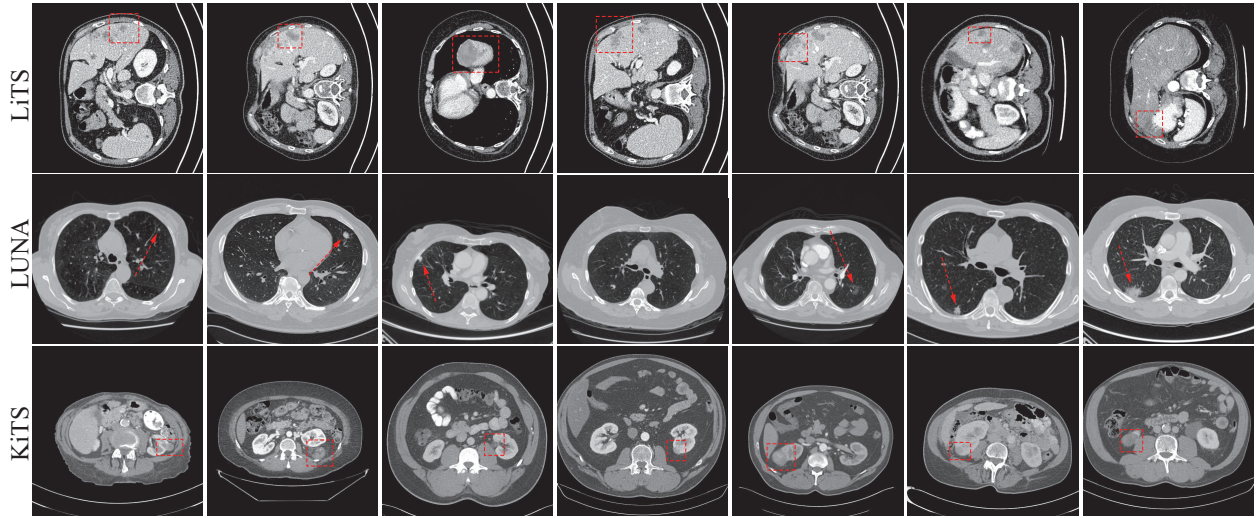}
  \caption{Synthetic tumors (red) generated by FRGAN on liver, lung, and kidney CT images from the LiTS, LUNA, and KiTS test datasets.}
  \label{fig:synthe_mag}
\end{figure*}


More examples of synthetic tumors in livers, lungs, and kidneys CT images generated by FRGAN are illustrated in Figure~\ref{fig:synthe_mag}. The visual results show that there is no difference between synthetic and real tumors on test datasets, which is consistent with quantitative evaluation results. The results demonstrate that integrating the DG-encoder, the RC-decoder, and the hybrid loss function into FRGAN can significantly enhance the learning of boundary features and contextual information to improve the whole image synthesis performance.

\begin{table*}[b]
  \centering
  \begin{threeparttable}
    \caption{Segmentation performance of the ablation study using the KiTS dataset.}
    \label{table:ablation_study}
    \renewcommand\arraystretch{1.1}
    \renewcommand\tabcolsep{3.0pt}
    \begin{tabular}{ccccccc|c}
      \toprule
      \tabincell{c}{Conventional data                                                                                                \\ augmentation} & \tabincell{c}{Vanilla \\ conv} & \tabincell{c}{DG-encoder} & RC-decoder & $L_{\mathrm{mm}}$ & $L_{\mathrm{percep}}$   & $L_{\mathrm{sty}}$  & Jaccard     \\ \hline
       &                      &                      &                      &                      &  &  & 0.737 \\
      \boldsymbol{$\surd$} &                      &                      &                      &                      &  &  & 0.752 \\
                           & \boldsymbol{$\surd$} &                      &                      &                      &  &  & 0.755 \\
                           &                      & \boldsymbol{$\surd$} &                      &
                           &
                           &
                           & 0.759
      \\
                           &                      & \boldsymbol{$\surd$} & \boldsymbol{$\surd$} &
                           &
                           &
                           & 0.758
      \\
                           &                      & \boldsymbol{$\surd$} & \boldsymbol{$\surd$} & \boldsymbol{$\surd$}
                           &
                           &
                           &0.760     \\
                           &                      & \boldsymbol{$\surd$} & \boldsymbol{$\surd$} &
                           & \boldsymbol{$\surd$}
                           &
                           & 0.762

      \\
                           &                      & \boldsymbol{$\surd$} & \boldsymbol{$\surd$} &
                           &
                           & \boldsymbol{$\surd$}
                           & 0.760
      \\
                           &                      & \boldsymbol{$\surd$} & \boldsymbol{$\surd$} & \boldsymbol{$\surd$}
                           & \boldsymbol{$\surd$}
                           & \boldsymbol{$\surd$}
                           & 0.763
      \\
      \bottomrule
    \end{tabular}
  \end{threeparttable}
\end{table*}

\subsection{Ablation studies}
To further evaluate and demonstrate the contributions of the major components in the proposed FRGAN model, we conduct ablation studies using the KiTS dataset. The newly introduced components include a DG-encoder, an RC-decoder, a multi-mask loss, a perceptual loss, and a style loss.

\subsubsection{Effectiveness of major components}
Ablation study is performed by gradually adding an individual component to investigate the importance of this newly added component. For quantitative evaluation, we also train a U-Net using the same real data together with an extra 20\% data generated by conventional augmentation of flipping and rotation. As shown in Table~\ref{table:ablation_study}, when using conventional data augmentation, the Jaccard score was improved by 0.5\% compared to the baseline model without augmentation. The experimental results showed that the spatial overlap in terms of Jaccard was improved with the DG-encoder. When the DG-encoder was used together with the RC-decoder as a complete generator, the evaluation results slightly dropped to Jaccard of 0.758. Although the performance dropped, the effectiveness of the RC-decoder is obvious as shown in Figure~\ref{fig:synthesis_ablation_results_add}. As shown in this figure, the tumor region synthesized by the model with the RC-decoder tended to be blurred and smoothed when compared to the previous synthesis model without the RC-decoder. With the RC-decoder, however, the tumor center and boundary regions became much more discriminative. In other words, the RC-decoder was able to reconstruct texture appearance within the tumor and near the tumor boundary. 

\subsubsection{Effectiveness of loss functions}
\begin{figure*}
  \centering
  \includegraphics[scale=1.0]{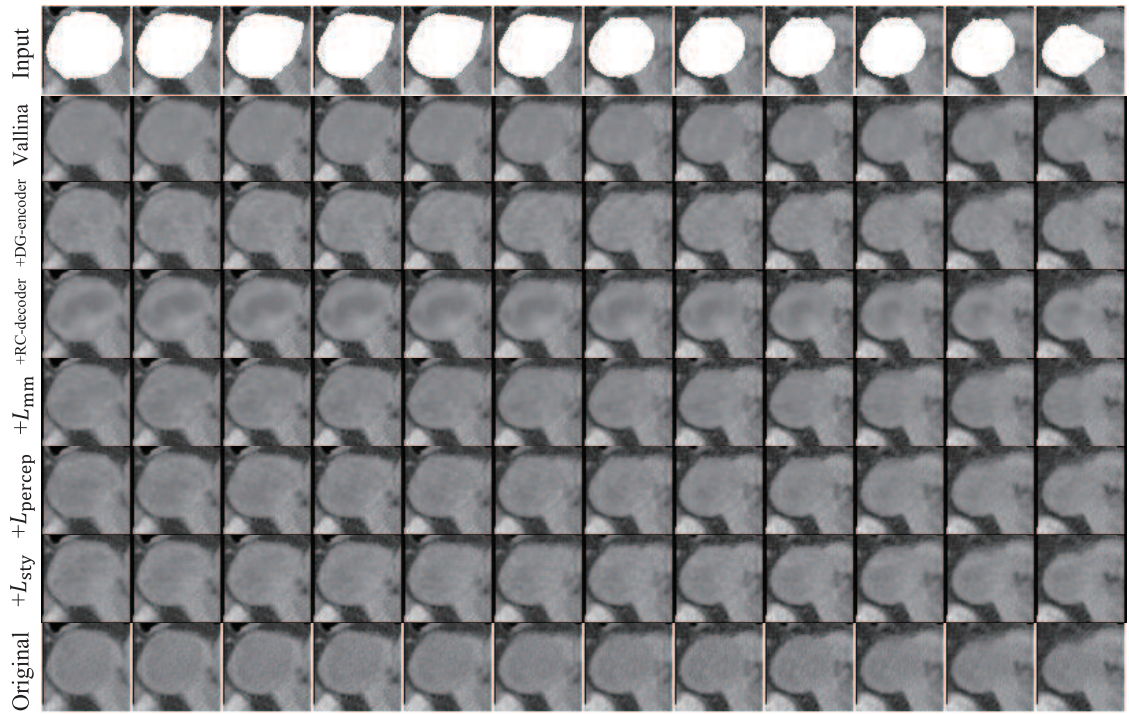}
  \caption{Synthesis results of ablation study on a 3D data sample of the KiTS validation dataset. Continuous slices in z-axis are flattened into 2D view from left column to right column for a better illustration. Each component of our model is gradually added from the top row to the bottom row.}
  \label{fig:synthesis_ablation_results_add}
\end{figure*}

In order to penalize the model on the tumor region and generate more realistic texture appearance, we added the multi-mask loss. As we can see from Table~\ref{table:ablation_study} and Figure~\ref{fig:synthesis_ablation_results_add}, the performance finally improved and the models could generate tumors with reasonable semantics and richer pathological textures. When we individually added the perceptual loss and the style loss, the segmentation performance improved simultaneously. All the losses in the hybrid loss function contributed to the synthesis results as the results were finally improved by adding the multi-mask loss, the perceptual loss, and the style loss.

The visual comparison results are given in Figure~\ref{fig:synthesis_ablation_results_add}. It can be seen that the synthesis results using vanilla convolution obtained the worst results where all the synthesis tumors were smoothed without texture information. In addition, the tumors were not properly fused to the surrounding tissues. When using the proposed generator with the DG-encoder, more realistic tumors are generated. The discriminant between tumor and boundaries are enhanced by the RC-decoder. However, blurriness is introduced by the RC-decoder as well. This is because the recovered features are only penalized and optimized in the synthesis process by the conventional loss function. When using the hybrid loss (i.e., $L_{\mathrm{mm}}$, $L_{\mathrm{percep}}$, and $L_{\mathrm{sty}}$) function to integrate the recovered information, as shown by the 5-8 rows in Figure~\ref{fig:synthesis_ablation_results_add}, apparently improved synthesizing results were obtained. The synthesis tumors were with texture features and fused with surrounding tissues naturally.

To this end, we draw the conclusions as below: (1) it clearly demonstrates that all the loss functions and the modules play important roles in determining the performance; (2) the style loss is not as effective as the perceptual loss in generating richer textures; and (3) the textural and spatial consistency are captured and reconstructed by FRGAN.

\subsection{Discussion}
\label{sec:discussion}
Our main findings are that the proposed tumor synthesis model: a) was able to generate synthesis tumors which were the closest to real tumors when compared with other models, and b) the synthesizing results improved the segmentation accuracy.

We explain the first finding by the dilated-gated generator with richer convolution providing our model with the capacity to extract and recover multi-scale features from the enlarged perceptive field. In addition, the multi-mask loss function increases the textural consistency among tumor, boundary, and background. Meanwhile, the perceptual and style losses are responsible for style transfer. Finally, the proposed hybrid loss function preserves the texture information during the synthesis process, leading to decreased blurriness. For instance, in the cases shown in Figure~\ref{fig:synthesis_ablation_results_add}, using the conventional loss function or vanilla convolution, the synthesizing results were smoothed without detailed content information. This is also the case with the two methods in comparison as shown in Figure~\ref{fig:synthe_comp}.

The first finding was also supported by Table~\ref{table:segmentation_performance}, where the models trained using our synthesis results generally outperformed the other methods in comparison, in terms of volumetric spatial overlap by Dice, Jaccard, VOE, and shape similarity by RVD and HD. This is because the segmentation model extracts and learns the characteristics from the samples in the training dataset. The testing results are obtained by applying a trained model to unseen real tumors collected from medical images. Given the same segmentation model and experimental settings, a training dataset composed of a greater number of diverse ``real'' tumor samples would have high probability leading to better testing results on real tumors. Thus, it can be inferred that the synthetic tumors/lesions generated by our model are the closest to the real ones among all the methods in comparison.

The second finding was supported by the comparison between models trained using real images alone, datasets of real images and traditional data augmentations, and datasets composed of real images and synthesis images in Tables~\ref{table:segmentation_performance} and~\ref{table:ablation_study}. According to our experimental results, all the segmentation models trained using synthesis images achieved improved performance than data augmentation or without data augmentation. This is because the real and synthetic training dataset is composed of tumor samples with diverse sizes, shapes, and locations. Thus, synthesis models enable segmentation models to utilize existing annotation effectively and provide an alternative solution to learn from a limited quantity of annotation.

One limitation of our method is that we assume the users, who can design appropriate tumor masks, are specialists with pathological knowledge or under the supervision of specialists with domain knowledge. Tumors, however, can appear anywhere in organs and have various types of shapes, sizes, and locations~\cite{HAVAEI201718}. The different sizes, shapes, and locations of the tumors could affect the complications and the malignancy rate of lesions~\cite{VENKATESH20061169}. Our future work includes proposing a new model with the capacity to force the discriminator to correct inappropriate input masks and generate tumors with pathological meanings.

\section{Conclusion}
\label{sec:conclusion}

We present a new user-customized 3D tumor synthesizing model for medical images. Our model leverages 3D gated convolution, richer convolutional feature enhanced dilated-gated generator, patch-based discriminator, and a hybrid loss function in a generative adversarial model. The evaluations conducted on liver CT, kidney CT, and lung nodule CT images using masks of different sizes, locations, and shapes demonstrate improved synthesizing results when compared with state-of-the-art methods. The proposed model has great potential to be applied to other modalities of medical images, especially when there are needs of generating user-customized tumors, soft tissues, or lesions in different medical image processing tasks. It may also assist medical training or surgeons to investigate novel cure for different cancers.

\section*{Acknowledgment}
This work is supported by the National Natural Science Foundation of China [Grant No. 61702361], the Science and Technology Program of Tianjin, China [Grant No. 16ZXHLGX00170], the National Key Technology R\&D Program of China [Grant No. 2018YFB1701700], and the program of China Scholarships Council.

\bibliography{elsarticle-template}

\end{document}